\documentclass[12pt]{iopart}
\usepackage{graphicx}
\usepackage{graphics,amssymb,bm}

\def\ket#1{|\,#1\,\rangle}

\def\braket#1#2{\langle\, #1\,|\,#2\,\rangle}

\def\ol#1{\overline{#1}}

\newcommand{\opx}{\hat{\sigma}_x}

\newcommand{\opz}{\hat{\sigma}_z}
\newcommand{\identity}{\hat I}

\newcommand{\beq}{\begin{equation}}
\newcommand{\eeq}{\end{equation}}
\newcommand{\beqa}{\begin{eqnarray}}
\newcommand{\eeqa}{\end{eqnarray}}

\begin{document}

\title[Generation of all sets of mutually unbiased bases]{Generation of all sets of mutually unbiased bases for three-qubit systems}

\author{Iulia Ghiu}

\address{Centre for Advanced Quantum Physics,
Department of Physics, University of Bucharest,
P. O.  Box MG-11, R-077125  Bucharest-M\u{a}gurele, Romania}

\ead{iulia.ghiu@g.unibuc.ro}

\begin{abstract}
We propose a new method of finding the mutually unbiased bases for three qubits. The key element is the construction of the table of striation-generating curves in the discrete phase space.
We derive a system of equations in the Galois field GF(8) and show that the solutions of these equations are sufficient for the construction of the general sets of complete mutually unbiased bases. A few examples are presented in order to show how our algorithm works in the cases: striation table with three, two axes, and one and no axis in the discrete phase space.
\end{abstract}

\maketitle

\section{Introduction}
The mutually unbiased bases (MUBs) represent a basic tool in many applications of quantum information processing: quantum tomography \cite{Klimov-2008}, quantum key distribution required in cryptography \cite{cripto}, discrete Wigner function \cite{Wootters-1987}, quantum teleportation \cite{telep}, quantum error correction codes \cite{err}, or mean king's problem \cite{Hayashi}.  Two bases $\{\ket{\psi_j}\}$ and $\{\ket{\phi_k}\}$ are called mutually unbiased if
$$|\braket{\psi_j}{\phi_k}|^2=\frac{1}{d}.$$
where $d$ is the dimension of the Hilbert space \cite{Wootters-1989}.

If $d$ is a power of a prime number, then there is method for finding the MUBs based on classes of commuting operators. One needs to determine $d + 1$ classes of $d - 1$ commuting operators, whose eigenvectors represent the MUBs. These special operators are called mutually unbiased operators \cite{Band}. The particular case of $N$-qubit systems was investigated by Lawrence {\it et al.}  \cite{Zeilinger}. The inseparability of the MUBs is discussed in Ref. \cite{Zeilinger} for two-qubit systems (one example) and for three-qubit systems (two examples).

MUBs can be constructed with the help of Galois fields GF($d$) in the special case when $d=p^n$ is the power of a prime number. Wootters \cite{Wootters-ibm} and Gibbons {\it et all.} \cite{Gibbons} proposed a method based on the so-called discrete phase space.
The discrete phase space of a d-level system is a $d \times d$ lattice, whose coordinates are elements of the finite Galois field GF($d$) \cite{Wootters-ibm}. A state of the MUBs is associated to a curve in the discrete phase space. The set of parallel curves is called a striation. There are $d+1$ striations. It turns out that the MUBs are determined by the states associated with each striation.

The construction of the general MUBs for two-qubit systems was found by Klimov {\it et all.} \cite{Klimov-2007}. In Ref. \cite{Bjork-2007}, Bj\"{o}rk {\it et al.} analyzed the MUB structures for three-qubit systems and obtained four structures:
(3,0,6), (2,3,4), (1,6,2), and (0,9,0), where the numbers represent
the number of triseparable, biseparable, and nonseparable bases,
respectively.

In this paper we generalize our method given in Ref. \cite{Ghiu} of generation of MUBs from two qubits to three qubits. Our purpose is to obtain all the sets of MUBs for three-qubit systems.

The paper is organized as follows: In Section 2 we present some basic notations and definitions used in the Galois fields. The main result of the paper is given in Section 3, where we find a system of 12 equations in the Galois field GF(8). The solution of these equations leads to the construction of the most general set of MUBs for three-qubit systems. Section 4 discusses the four algorithms needed for obtaining the striation-generating curves in the cases: striation table with three, two axes, and one and no axis in the discrete phase space. One example for each algorithm is given in Section 5.
Finally, we make some concluding remarks in Section 6.

\section{Preliminaries on Galois fields}

In this section we recall some definitions and notations used in Galois fields.
The map trace of a field element $x \in $ GF($p^n$) is as follows:
$$\mbox{tr}\; x =x +x^2 +... +x^{p^{n-1}}. $$

If we denote by $\mu $ the primitive element of GF(8), then the eight elements of GF(8) are $\{0, 1, \mu , \mu^2, \mu^3, \mu^4, \mu^5, \mu^6\}$. The irreducible polynomial is: $\mu ^3+\mu +1$. A basis $\{ \alpha_i\}$ is called self-dual if $\mbox{tr}  (\alpha_i \alpha_j)  = \delta_{ij}$. We choose the self-dual basis to be $\{ \mu^3, \mu^5, \mu^6\}$.

Following Ref. \cite{Bjork-2007}, we associate the points in the phase space which contain the self-dual basis elements with the following operators:
\begin{eqnarray}
(\mu^3,0) & \leftrightarrow & \opx \otimes \identity \otimes \identity ,
\nonumber \\
(\mu^5,0) & \leftrightarrow & \identity \otimes \opx \otimes \identity , \nonumber \\
(\mu^6,0) & \leftrightarrow & \identity \otimes \identity \otimes \opx ,
\nonumber \\
(0,\mu^3) & \leftrightarrow & \opz \otimes \identity \otimes \identity ,
\nonumber \\
(0,\mu^5) & \leftrightarrow & \identity \otimes \opz \otimes \identity , \nonumber \\
(0,\mu^6) & \leftrightarrow & \identity \otimes \identity \otimes \opz .
\nonumber
\end{eqnarray}

\section{The system of 12 equations in the Galois field GF(8)}

Two operators commute if their associated points in the discrete phase space $(a_1,b_1)$ and $(a_2,b_2)$ satisfy \cite{Klimov-2009}
\beq
\mbox{tr}(a_1b_2)=\mbox{tr}(a_2b_1).
\label{rel-com-gen}
\eeq

The MUBs can be constructed with the help of $d+1$ classes of $d-1$ commuting operators, i.e. MU operators. In the case of three-qubit systems, we need to generate a table with 9 rows, each row representing the set of 7 commuting operators. We denote the operator which belongs to the row $r$ and column $c$ by $O_{r,c}$, where $r=\ol{1,9}$ and $c=\ol{1,7}$.
The point in the phase space $(a^{(r)}_c,b^{(r)}_c)$ is associated to the operator $O_{r,c}$:
$$ O_{r,c}\longleftrightarrow   (a^{(r)}_c,b^{(r)}_c).$$

There are six operators which define uniquely a table of nine sets of operators that generate a certain MUB structure \cite{Bjork-2007}. The operators in the first two rows are defined
by $\hat{O}_{r,c} = \hat{O}_{r,c-2} \hat{O}_{r,c-3}$, where $r=
1,2$ and $c = 1, \ldots , 7$. The other rows are obtained as $\hat{O}_{r,c} =
\hat{O}_{2,c} \hat{O}_{1,c+r-3}$, for $2 < r \leq 9$ \cite{Bjork-2007}.

We denote the six points in the discrete phase space associated to these six operators by:
\beqa \label{input-op}
&&(a^{(1)}_1,b^{(1)}_1); \;  (a^{(1)}_2,b^{(1)}_2); \; (a^{(1)}_3,b^{(1)}_3);\nonumber\\
&&(a^{(2)}_1,b^{(2)}_1); \;  (a^{(2)}_2,b^{(2)}_2); \; (a^{(2)}_3,b^{(2)}_3).
\eeqa

We want to find the conditions that should be fulfilled by these 6 points, i.e. 12 variables, in order to obtain a table of striation-generating curves, that generates MUBs.

We define
\beq \label{def1}
(a^{(r)}_c,b^{(r)}_c):=(a^{(r)}_{c-2}+a^{(r)}_{c-3},b^{(r)}_{c-2}+b^{(r)}_{c-3})
\eeq
for $r=1,2$ and $c=4,5,6,7$ and
\beq \label{def2}
(a^{(r)}_c,b^{(r)}_c):=(a^{(2)}_c+a^{(1)}_{c+r-3},b^{(2)}_c+b^{(1)}_{c+r-3})
\eeq
for $r=\ol{3,9}$ and $c=\ol{1,7}$ and the sum of the indices is taken modulo 7.

We have to impose the condition (\ref{rel-com-gen}) for all operators of striation 1 and 2, respectively:
\beqa \label{str12}
tr\left[ a^{(1)}_jb^{(1)}_k\right]&=&tr\left[ a^{(1)}_kb^{(1)}_j\right],\nonumber\\
tr\left[ a^{(2)}_jb^{(2)}_k\right]&=&tr\left[ a^{(2)}_kb^{(2)}_j\right] ,
\eeqa
where $j\not= k=\ol{1,7}$.
A striation-generating curve is well defined by three points, although its parametric form is not unique. Therefore it is sufficient to check the condition (\ref{rel-com-gen}) only for the first three points of the striations 3,...,9.

The first three points of the striations 3,..., 9 are obtained from Eq. (\ref{def2}).
For a fixed striation $r$, we have the following conditions:
\beqa \label{com39new}
&&tr\left[ a^{(2)}_1b^{(1)}_{r-1}+a^{(1)}_{r-2}b^{(2)}_2\right] =tr\left[ a^{(2)}_2b^{(1)}_{r-2}+a^{(1)}_{r-1}b^{(2)}_1\right] ;\nonumber\\
&&tr\left[ a^{(2)}_1b^{(1)}_r+a^{(1)}_{r-2}b^{(2)}_3\right] =tr\left[ a^{(2)}_3b^{(1)}_{r-2}+a^{(1)}_rb^{(2)}_1\right] ;\nonumber\\
&&tr\left[ a^{(2)}_2b^{(1)}_r+a^{(1)}_{r-1}b^{(2)}_3\right] =tr\left[ a^{(2)}_3b^{(1)}_{r-1}+a^{(1)}_rb^{(2)}_2\right] .
\eeqa

By employing the definition (\ref{def2}), we notice that all the first three points of the striations $\ol{3,9}$ are expressed in terms of the six points of Eq. (\ref{input-op}). Therefore by writing the condition (\ref{rel-com-gen}) for them, we obtain 21 equations which contain the 12 parameters: $a^{(\lambda )}_j,b^{(\lambda )}_j$, where $j=1,2,3$ and $\lambda =1,2$ of the input points.

We proved that the 21 equations are not independent, i.e. the conditions (\ref{str12}) which correspond to striations 1 and 2 and Eq. (\ref{com39new}) written for striations 3 and 4 will generate the other 15 equations. In conclusion, it is sufficient that the condition (\ref{rel-com-gen}) to be satisfied for the first three points of striations 1, 2, 3, and 4 for obtaining a striation table.

We write explicitly the 12 independent equations:
\beqa
&&tr\left[ a^{(1)}_1b^{(1)}_2\right]=tr\left[ a^{(1)}_2b^{(1)}_1\right]; \label{ec22}\\
&&tr\left[ a^{(1)}_1b^{(1)}_3\right]=tr\left[ a^{(1)}_3b^{(1)}_1\right]; \label{ec23}\\
&&tr\left[ a^{(1)}_2b^{(1)}_3\right]=tr\left[ a^{(1)}_3b^{(1)}_2\right]; \label{ec24}\\
&&tr\left[ a^{(2)}_1b^{(2)}_2\right]=tr\left[ a^{(2)}_2b^{(2)}_1\right]; \label{ec25}\\
&&tr\left[ a^{(2)}_1b^{(2)}_3\right]=tr\left[ a^{(2)}_3b^{(2)}_1\right]; \label{ec26}\\
&&tr\left[ a^{(2)}_2b^{(2)}_3\right]=tr\left[ a^{(2)}_3b^{(2)}_2\right]. \label{ec27}\\
&&tr\left[ a^{(2)}_1b^{(1)}_2+a^{(1)}_1b^{(2)}_2\right] =tr \left[ a^{(2)}_2b^{(1)}_1+a^{(1)}_2b^{(2)}_1\right]; \label{ec1}\\
&&tr\left[ a^{(2)}_1b^{(1)}_3+a^{(1)}_1b^{(2)}_3\right] =tr \left[ a^{(2)}_3b^{(1)}_1+a^{(1)}_3b^{(2)}_1\right]; \label{ec2}\\
&&tr\left[ a^{(2)}_2b^{(1)}_3+a^{(1)}_2b^{(2)}_3\right] =tr \left[ a^{(2)}_3b^{(1)}_2+a^{(1)}_3b^{(2)}_2\right]; \label{ec3}\\
&&tr\left[ a^{(2)}_1b^{(1)}_3+a^{(1)}_2b^{(2)}_2\right] =tr \left[ a^{(2)}_2b^{(1)}_2+a^{(1)}_3b^{(2)}_1\right]; \label{ec4}\\
&&tr\left[ a^{(2)}_1b^{(1)}_1+a^{(2)}_1b^{(1)}_2+a^{(1)}_2b^{(2)}_3\right] =tr \left[ a^{(2)}_3b^{(1)}_2+a^{(1)}_1b^{(2)}_1+a^{(1)}_2b^{(2)}_1\right]; \label{ec5}\\
&&tr\left[ a^{(2)}_2b^{(1)}_1+a^{(2)}_2b^{(1)}_2+a^{(1)}_3b^{(2)}_3\right] =tr \left[ a^{(2)}_3b^{(1)}_3+a^{(1)}_1b^{(2)}_2+a^{(1)}_2b^{(2)}_2\right] .\label{ec6}
\eeqa

In conclusion, we have to find the 12 parameters $a^{(1)}_1$, $b^{(1)}_1$, $a^{(1)}_2$, $b^{(1)}_2$, $a^{(1)}_3$, $b^{(1)}_3$, $a^{(2)}_1$, $b^{(2)}_1$, $a^{(2)}_2$, $b^{(2)}_2$, $a^{(2)}_3$, $b^{(2)}_3$ which satisfy the 12 equations  (\ref{ec22})$-$(\ref{ec6}). Having obtained them, we generate the whole table of striations by employing the definitions (\ref{def1}) and (\ref{def2}).

\section{The table of striation-generating curves of the general sets of mutually unbiased bases for three-qubit systems}

\subsection{The striation table with three axes in the discrete phase space}
The six points which define the striation table are:
\beqa
&&(0,\lambda_1); \;  (0,\lambda_2); \; (0,\lambda_3);\nonumber\\
&&(\lambda_1,0); \;  (\lambda_2,0); \; (\lambda_3,0).
\eeqa
The 12 equations (\ref{ec22})$-$(\ref{ec6}) are equivalent to the following three ones:
\beqa
&&tr\lambda _3=tr(\lambda_2+\lambda_1\lambda_2)\label{ec46}\\
&&tr(\lambda_1\lambda_3)=tr\lambda_2\label{ec44}\\
&&tr(\lambda_2\lambda_3)=tr(\lambda_1+\lambda_1\lambda_2).\label{ec45}
\eeqa

The algorithm for obtaining the six points which generate the striation table is:
\begin{itemize}
\item We fix $\lambda_1$ and $\lambda_2$, which have to be different.
\item The parameter $\lambda_3$ is found such that it satisfies the Eqs. (\ref{ec46})$-$(\ref{ec45}).
\end{itemize}

\subsection{The striation table with two axes in the discrete phase space}

The six points which define the striation table are:
\beqa
&&(0,b^{(1)}_1); \;  (0,b^{(1)}_2); \; (0,b^{(1)}_3);\nonumber\\
&&(a^{(2)}_1,0); \;  (a^{(2)}_2,0); \; (a^{(2)}_3,0).
\eeqa

The algorithm requires fixing 4 parameters, namely $b^{(1)}_1, b^{(1)}_2, b^{(1)}_3$, and $a^{(2)}_1$, and is as follows:
\begin{itemize}
\item The parameters $b^{(1)}_1, b^{(1)}_2, b^{(1)}_3$ are chosen such that they form a basis.
\item The parameter $a^{(2)}_1$ is arbitrary.
\item The other two parameters $a^{(2)}_2$ and $a^{(2)}_3$ are obtained from the Eqs. (\ref{ec1})$-$(\ref{ec6}).
\end{itemize}

\subsection{The striation table with one axis in the discrete phase space}
In this case the 6 points have the form:
\beqa
&&(0,b^{(1)}_1); \;  (0,b^{(1)}_2); \; (0,b^{(1)}_3);\nonumber\\
&&(a^{(2)}_1,b^{(2)}_1); \;  (a^{(2)}_2,b^{(2)}_2); \; (a^{(2)}_3,b^{(2)}_3).
\eeqa

The algorithm is as follows:
\begin{itemize}
\item The parameters $b^{(1)}_1, b^{(1)}_2, b^{(1)}_3$ are chosen such that they form a basis.
\item The parameters $a^{(2)}_1, b^{(2)}_2,b^{(2)}_3$ are arbitrary.
\item The remaining 3 parameters $b^{(2)}_1,a^{(2)}_2,a^{(2)}_3$ are obtained from the Eqs. (\ref{ec25})$-$(\ref{ec6}).
\end{itemize}

\subsection{The striation table with no axis in the discrete phase space}

The six points which define the striation table with no axis are:
\beqa
&&(a^{(1)}_1,b^{(1)}_1); \;  (a^{(1)}_2,b^{(1)}_2); \; (a^{(1)}_3,b^{(1)}_3);\nonumber\\
&&(a^{(2)}_1,b^{(2)}_1); \;  (a^{(2)}_2,b^{(2)}_2); \; (a^{(2)}_3,b^{(2)}_3).
\eeqa

The algorithm requires in this case to fix 7 parameters, namely $a^{(1)}_1$, $b^{(1)}_1$, $b^{(1)}_2$, $b^{(1)}_3$, $a^{(2)}_1$, $b^{(2)}_2$, $b^{(2)}_3$:
\begin{itemize}
\item We have to fix the parameters $a^{(1)}_1, b^{(1)}_1, b^{(1)}_2, b^{(1)}_3, a^{(2)}_1, b^{(2)}_2,b^{(2)}_3$.
\item The other 5 parameters $a^{(2)}_1,a^{(1)}_3,b^{(2)}_1,a^{(2)}_2,a^{(2)}_3$ are found such that they verify all the Eqs. (\ref{ec22})$-$(\ref{ec6}).
\end{itemize}

\section{Examples}

\subsection{The striation table with three axes}

We take $\lambda_1=\mu^2$ and $\lambda_2=\mu^6$. The Eqs. (\ref{ec46})$-$(\ref{ec45}) are:
$$tr\lambda_3=1; \; tr(\mu^2\lambda_3)=1; \; tr(\mu^6\lambda_3)=0.
$$
We find two solutions: $\lambda_3=\mu^3$ and $\lambda_3=\mu^5$. The striation tables are identical to the two tables shown in Fig. 3 of Ref. \cite{Bjork-2007}. The expressions of the curves for the first solution $\lambda_3=\mu^3$ are given in Table \ref{tabel-3axe}. We denote by $(\alpha ,\beta )$ a point in the discrete phase space.

\begin{table}
\begin{center}
\caption{The striation-generating curves corresponding to the table with three axes: the solution with $\lambda_3=\mu^3$.}
\begin{tabular}{c||c|c|c|c|c|c|c|c|}
\hline
$\mu^6$&$\,${\footnotesize 1}$\,$ &$\,${\footnotesize 8}$\,$&$\,${\footnotesize 6}$\,$&$\,${\footnotesize 4}$\,$&$\,${\footnotesize 9}$\,$&$\,${\footnotesize 7}$\,$&$\,${\footnotesize 5}$\,$&$\,${\footnotesize 3}$\,$\\ \hline
$\mu^5$& {\footnotesize 1} & {\footnotesize 6} & {\footnotesize 4} & {\footnotesize 9} & {\footnotesize 7} & {\footnotesize 5} & {\footnotesize 3} & {\footnotesize 8}\\ \hline
$\mu^4$& {\footnotesize 1} & {\footnotesize 4} & {\footnotesize 9} & {\footnotesize 7} & {\footnotesize 5} & {\footnotesize 3} & {\footnotesize 8} & {\footnotesize 6}\\ \hline
$\mu^3$& {\footnotesize 1} & {\footnotesize 9} & {\footnotesize 7} & {\footnotesize 5} & {\footnotesize 3} & {\footnotesize 8} & {\footnotesize 6} & {\footnotesize 4}\\ \hline
$\mu^2$& {\footnotesize 1} & {\footnotesize 7} & {\footnotesize 5} & {\footnotesize 3} & {\footnotesize 8} & {\footnotesize 6} & {\footnotesize 4} & {\footnotesize 9}\\ \hline
$\mu $& {\footnotesize 1} & {\footnotesize 5} & {\footnotesize 3} & {\footnotesize 8} & {\footnotesize 6} & {\footnotesize 4} & {\footnotesize 9} & {\footnotesize 7}\\ \hline
1& {\footnotesize 1} & {\footnotesize 3} & {\footnotesize 8} & {\footnotesize 6} & {\footnotesize 4} & {\footnotesize 9} & {\footnotesize 7} & {\footnotesize 5}\\ \hline
0&{\footnotesize o} & {\footnotesize 2} & {\footnotesize 2} & {\footnotesize 2} & {\footnotesize 2} & {\footnotesize 2} & {\footnotesize 2} & {\footnotesize 2}\\
\hline   \hline      &$0$&$1$&$\mu$&$\mu^2$&$\mu^3$&$\mu^4$&$\mu^5$&$\mu^6$
\\
\end{tabular}
\hspace{0.5cm}
\begin{tabular}{|c|c|}
\hline
No. & Equation \\ \hline
1& $\alpha $=0 and $\beta (k)=k$ \\ \hline
2& $\alpha (k)=k$ and $\beta =0$ \\ \hline
3& $\beta =\alpha $\\ \hline
4& $\beta =\alpha \mu $ \\ \hline
5& $\beta =\alpha \mu^2 $ \\ \hline
6& $\beta =\alpha \mu^3 $ \\ \hline
7& $\beta =\alpha \mu^4 $ \\ \hline
8& $\beta =\alpha \mu^5 $ \\ \hline
9& $\beta =\alpha \mu^6 $ \\ \hline
\end{tabular}
\label{tabel-3axe}
\end{center}
\end{table}

\subsection{The table with two axes}
We start with
\beqa
&&(0,\mu^4); \;  (0,\mu ^3); \; (0,\mu ^5);\nonumber\\
&&(1,0); \;  (a^{(2)}_2,0); \; (a^{(2)}_3,0).
\eeqa
The system of six equation is:
$$tr \left[ a^{(2)}_2\mu^4\right] =1; \; tr \left[ a^{(2)}_3\mu^4\right] =1; \;
tr \left[ a^{(2)}_3\mu^3\right] =tr\left[ a^{(2)}_2\mu^5\right].$$
$$tr \left[ a^{(2)}_2\mu^3\right] =1; \; tr \left[ a^{(2)}_3\mu^3\right] =1; \;
tr \left[ a^{(2)}_3\mu^5\right] =tr\left[ a^{(2)}_2\mu^6\right].$$

We obtain the unique solution $a^{(2)}_2=\mu^2$  and $a^{(2)}_3=\mu^3$.

The 9 generating striations and their analytical expressions are given in Table \ref{tabel-2axe}.

\begin{table}
\begin{center}
\caption{The striation-generating curves corresponding to the table with two axes.}
\begin{tabular}{|c|c|c|c|c|c|c|c|}
\hline
$\,${\footnotesize 1}$\,$ & $\,${\footnotesize 6}$\,$ & $\,${\footnotesize 7}$\,$& $\,${\footnotesize 5}$\,$& $\,${\footnotesize 4}$\,$ & $\,${\footnotesize 8}$\,$ & $\,${\footnotesize 9}$\,$ & $\,${\footnotesize 3}$\,$  \\  \hline
$\,${\footnotesize 1}$\,$&{\footnotesize 5}&{\footnotesize 6}&{\footnotesize 4}&{\footnotesize 3}&{\footnotesize 7}&{\footnotesize 8}&{\footnotesize 9}\\ \hline
$\,${\footnotesize 1}$\,$&{\footnotesize 3}&{\footnotesize 4}&{\footnotesize 9}&{\footnotesize 8}&{\footnotesize 5}&{\footnotesize 6}&{\footnotesize 7} \\ \hline
$\,${\footnotesize 1}$\,$&{\footnotesize 4}&{\footnotesize 5}&{\footnotesize 3}&{\footnotesize 9}&{\footnotesize 6}&{\footnotesize 7}&{\footnotesize 8}  \\ \hline
$\,${\footnotesize 1}$\,$&{\footnotesize 7}&{\footnotesize 8}&{\footnotesize 6}&{\footnotesize 5}&{\footnotesize 9}&{\footnotesize 3}&{\footnotesize 4} \\ \hline
$\,${\footnotesize 1}$\,$&{\footnotesize 8}&{\footnotesize 9}&{\footnotesize 7}&{\footnotesize 6}&{\footnotesize 3}&{\footnotesize 4}&{\footnotesize 5}  \\ \hline
$\,${\footnotesize 1}$\,$&{\footnotesize 9}&{\footnotesize 3}&{\footnotesize 8}&{\footnotesize 7}&{\footnotesize 4}&{\footnotesize 5}&{\footnotesize 6}  \\ \hline
$\,${\footnotesize o}$\,$&{\footnotesize 2}&{\footnotesize 2}&{\footnotesize 2}&{\footnotesize 2}&{\footnotesize 2}&{\footnotesize 2}&{\footnotesize 2}  \\ \hline
\end{tabular}
\hspace{0.5cm}
\begin{tabular}{|c|c|}
\hline
No. & Equation \\ \hline
1& $\alpha $=0 and $\beta (k)=k$ \\ \hline
2& $\alpha (k)=k$ and $\beta =0$ \\ \hline
3& $\beta =\mu^2\alpha +\mu^5\alpha^2+\mu^6\alpha^4 $\\ \hline
4& $\beta =\mu^3\alpha$\\ \hline
5& $\beta =\alpha +\mu^2\alpha^2+\mu \alpha^4$\\ \hline
6& $\beta =\mu^5\alpha +\mu^5\alpha^2+\mu ^6\alpha^4$\\ \hline
7& $\beta =\mu \alpha +\mu^2\alpha^2+\mu \alpha^4 $\\ \hline
8& $\beta =\mu^4\alpha +\mu^3\alpha^2+\mu ^5\alpha^4$\\ \hline
9& $\beta =\mu^6\alpha +\mu^3\alpha^2+\mu ^5\alpha^4$\\ \hline
\end{tabular}
\label{tabel-2axe}
\end{center}
\end{table}

\subsection{The striation table with one axis}
We start with
\beqa
&&(0,\mu^4); \;  (0,\mu^3); \; (0,\mu );\nonumber\\
&&(1,b^{(2)}_1); \;  (a^{(2)}_2,\mu^2); \; (a^{(2)}_3,\mu^6).
\eeqa
The system of 9 equations is:
$$tr\left[ a^{(2)}_2b^{(2)}_1\right]=0; \;
tr\left[ a^{(2)}_3b^{(2)}_1\right]=1; \;
tr\left[ a^{(2)}_2\mu^6\right]=tr\left[ a^{(2)}_3\mu^2\right]; $$
$$tr\left[ a^{(2)}_2\mu^4\right] =1; \;
tr\left[ a^{(2)}_3\mu^4\right]=0; \;
tr\left[ a^{(2)}_2\mu+ a^{(2)}_3\mu^3\right]=0; $$
$$tr\left[ a^{(2)}_2\mu^3\right] =0; \;
tr\left[ a^{(2)}_3\mu^3\right] =1; \;
tr\left[ a^{(2)}_2\mu^6+a^{(2)}_3\mu\right] =0.$$

We find 2 solutions:
\beqa
&& a^{(2)}_2=\mu^6,a^{(2)}_3=\mu^4,b^{(2)}_1=\mu^2 \hspace{0.5cm} \mbox{and}\nonumber\\
&& a^{(2)}_2=\mu^6,a^{(2)}_3=\mu^4,b^{(2)}_1=\mu^3.
\eeqa

The expressions of the curves for the first solution $b^{(2)}_1=\mu^2$ are shown in Table \ref{tabel-1axa}.

\begin{table}
\begin{center}
\caption{The striation-generating curves corresponding to the table with one axis.}
\begin{tabular}{|c|c|c|c|c|c|c|c|}
\hline
$\,${\footnotesize 1}$\,$ & $\,${\footnotesize 7}$\,$ & $\,${\footnotesize 2}$\,$ & $\,${\footnotesize 3}$\,$ & $\,${\footnotesize 5}$\,$ & $\,${\footnotesize 2}$\,$ & $\,${\footnotesize 3}$\,$ & $\,${\footnotesize 6}$\,$ \\  \hline
$\,${\footnotesize 1}$\,$ &{\footnotesize 4}&{\footnotesize 7}&{\footnotesize 5}&{\footnotesize 6}&{\footnotesize 3}&{\footnotesize 4}&{\footnotesize 3}\\ \hline
$\,${\footnotesize 1}$\,$ &{\footnotesize 5}&{\footnotesize 6}&{\footnotesize 7}&{\footnotesize 4}&{\footnotesize 9}&{\footnotesize 9}&{\footnotesize 4}\\ \hline
$\,${\footnotesize 1}$\,$ &{\footnotesize 8}&{\footnotesize 5}&{\footnotesize 8}&{\footnotesize 8}&{\footnotesize 8}&{\footnotesize 6}&{\footnotesize 7}\\ \hline
$\,${\footnotesize 1}$\,$ &{\footnotesize 2}&{\footnotesize 9}&{\footnotesize 6}&{\footnotesize 9}&{\footnotesize 5}&{\footnotesize 7}&{\footnotesize 2}\\ \hline
$\,${\footnotesize 1}$\,$ &{\footnotesize 3}&{\footnotesize 3}&{\footnotesize 9}&{\footnotesize 7}&{\footnotesize 6}&{\footnotesize 5}&{\footnotesize 9}\\ \hline
$\,${\footnotesize 1}$\,$ &{\footnotesize 6}&{\footnotesize 4}&{\footnotesize 4}&{\footnotesize 2}&{\footnotesize 7}&{\footnotesize 2}&{\footnotesize 5}\\ \hline
$\,${\footnotesize o}$\,$ &{\footnotesize 9}&{\footnotesize 8}&{\footnotesize 2}&{\footnotesize 3}&{\footnotesize 4}&{\footnotesize 8}&{\footnotesize 8}\\ \hline
\end{tabular}
\hspace{0.5cm}
\begin{tabular}{|c|c|}
\hline
No. & Equation \\ \hline
1& $\alpha $=0 and $\beta (k)=k$\\ \hline
2& $\beta =\mu ^5\alpha +\alpha^2+\mu^3\alpha^4$\\ \hline
3& $\beta =\mu^5\alpha ^2+\mu^6\alpha^4 $ \\ \hline
4& $\beta =\alpha +\mu ^6\alpha^2+\mu^3\alpha^4$\\ \hline
5& $\beta =\mu ^2\alpha +\mu^4\alpha^2+\mu ^2\alpha^4$\\ \hline
6& $\beta =\mu^5 \alpha +\mu ^6\alpha^2+\mu ^3\alpha^4$\\ \hline
7& $\beta =\mu ^5\alpha +\mu^5\alpha^2+\mu ^6\alpha^4$\\ \hline
8& $\beta =\mu^6\alpha +\mu^2\alpha^2+\mu \alpha^4$\\ \hline
9& $\beta =\mu \alpha +\mu^4\alpha^2+\mu ^2\alpha^4$.\\ \hline
\end{tabular}
\label{tabel-1axa}
\end{center}
\end{table}

\subsection{The striation table with no axis}
We start with the following points:
\beqa
&&(\mu^2,\mu^5); \;  (a^{(1)}_2,\mu^3); \; (a^{(1)}_3,1);\nonumber\\
&&(\mu^3,b^{(2)}_1); \;  (a^{(2)}_2,\mu^2); \; (a^{(2)}_3,\mu).
\eeqa

The 12 equations are:
$$tr\left[ a^{(1)}_2\mu^5\right]=1; \;
tr\left[ a^{(1)}_3\mu^5\right]=0; \;
tr \, a^{(1)}_2=tr\left[ a^{(1)}_3\mu^3\right]; \;
tr\left[ a^{(2)}_2b^{(2)}_1\right]=1; $$
$$tr\left[ a^{(2)}_3b^{(2)}_1\right]=0; \;
tr\left[ a^{(2)}_2\mu \right]=tr\left[ a^{(2)}_3\mu^2\right]; \;
tr \left[ a^{(2)}_2\mu^5+a^{(1)}_2b^{(2)}_1\right]=1; \;
tr \left[ a^{(2)}_3\mu^5+a^{(1)}_3b^{(2)}_1\right]=0; $$
$$tr\left[ a^{(2)}_2+a^{(1)}_2\mu +a^{(2)}_3\mu^3 +a^{(1)}_3\mu^2\right]=0; \;
tr\left[ a^{(1)}_2\mu^2+a^{(2)}_2\mu^3+a^{(1)}_3b^{(2)}_1\right] =1;$$
$$tr \left[ a^{(1)}_2\mu +a^{(2)}_3\mu^3 +a^{(1)}_2b^{(2)}_1\right]=tr\left[ \mu^5+\mu^2b^{(2)}_1\right]; \;
tr\left[ a^{(2)}_2\mu^2+a^{(1)}_3\mu +a^{(2)}_3+a^{(1)}_2\mu^2\right] =0.$$

One solution is the following:
$$a^{(1)}_2=1; \; a^{(1)}_3=\mu^3; \; a^{(2)}_2=\mu ; \; a^{(2)}_3=1; \; b^{(2)}_1=\mu^2 .$$

The table of striation-generating curves and their expressions are given in Table \ref{tabel-nicioaxa}.

\begin{table}
\begin{center}
\caption{The striation-generating curves corresponding to the table with no axis.}
\begin{tabular}{|c|c|c|c|c|c|c|c|}
\hline
$\,${\footnotesize 5}$\,$ & $\,${\footnotesize 4}$\,$ & $\,${\footnotesize 7}$\,$ & $\,${\footnotesize 5}$\,$ & $\,${\footnotesize 5}$\,$ & $\,${\footnotesize 1}$\,$ & $\,${\footnotesize 5}$\,$ & $\,${\footnotesize 8}$\,$ \\  \hline
$\,${\footnotesize 8}$\,$&{\footnotesize 6}&{\footnotesize 4}&{\footnotesize 1}&{\footnotesize 3}&{\footnotesize 3}&{\footnotesize 6}&{\footnotesize 7} \\ \hline
$\,${\footnotesize 6}$\,$&{\footnotesize 7}&{\footnotesize 2}&{\footnotesize 4}&{\footnotesize 2}&{\footnotesize 6}&{\footnotesize 1}&{\footnotesize 8}\\ \hline
$\,${\footnotesize 8}$\,$&{\footnotesize 1}&{\footnotesize 3}&{\footnotesize 9}&{\footnotesize 7}&{\footnotesize 9}&{\footnotesize 3}&{\footnotesize 4}\\ \hline
$\,${\footnotesize 8}$\,$&{\footnotesize 3}&{\footnotesize 2}&{\footnotesize 3}&{\footnotesize 2}&{\footnotesize 7}&{\footnotesize 4}&{\footnotesize 1}\\ \hline
$\,${\footnotesize 2}$\,$&{\footnotesize 2}&{\footnotesize 1}&{\footnotesize 9}&{\footnotesize 4}&{\footnotesize 9}&{\footnotesize 7}&{\footnotesize 8}\\ \hline
$\,${\footnotesize 9}$\,$&{\footnotesize 6}&{\footnotesize 9}&{\footnotesize 7}&{\footnotesize 1}&{\footnotesize 4}&{\footnotesize 6}&{\footnotesize 8}\\ \hline
$\,${\footnotesize o}$\,$&{\footnotesize 2}&{\footnotesize 9}&{\footnotesize 5}&{\footnotesize 5}&{\footnotesize 6}&{\footnotesize 5}&{\footnotesize 3}\\ \hline
\end{tabular}
\hspace{0.5cm}
\begin{tabular}{|c|c|}
\hline
No. & Equation \\ \hline
1& $\beta =\mu^6\alpha +\mu^6\alpha^2+\mu^3\alpha^4$\\ \hline
2& $\beta^2+\mu \beta = \mu^2\alpha ^2+\mu^2\alpha $\\ \hline
3& $\beta =\mu^3\alpha ^2+\mu^5\alpha^4 $\\ \hline
4& $\beta =\mu ^3\alpha +\mu ^2\alpha^2+\mu \alpha^4$\\ \hline
5& $\mu^3\beta^2+\mu^2\beta =\alpha^4+\mu^2\alpha^2+\mu^3\alpha $\\ \hline
6& $\mu \beta^2+\mu^5\beta =\alpha^4+\mu^5\alpha^2+\mu^3\alpha $\\ \hline
7& $\beta =\mu^6\alpha +\alpha^2+\alpha^4$\\ \hline
8& $\alpha =\mu^2\beta ^4+\mu^4\beta^2+\mu^5\beta $\\ \hline
9& $\mu \beta^2+\mu \beta =\alpha^4+\mu^2\alpha^2 $\\ \hline
\end{tabular}
\label{tabel-nicioaxa}
\end{center}
\end{table}

\section{Conclusions}
In this paper we have presented a method of construction of all sets of MUBs for three-qubit systems. The main result is the system of 12 equations (\ref{ec22})$-$(\ref{ec6}) in the Galois field GF(8). The solutions of this system are sufficient for the construction of different sets of MUBs for three qubits.

In the case of $N$ qubits, one needs to determine $2^N+1$ classes of $2^N-1$ commuting operators. Our method can be generalized to the case $N>3$, but the system of equations in the Galois field GF($2^N$) is much more complex, since the number of equations increases.

\ack
I wish to thank Gunnar Bj\"{o}rk, Andrei B. Klimov, Luis L. S\'anchez-Soto, and Cristian Ghiu for useful discussions on mutually unbiased bases and Galois fields. This work was supported by CNCS - UEFISCSU, postdoctoral research project PD code 151, no. 150/30.07.2010 for the University of Bucharest.


\section*{References}

\begin{thebibliography}{99}

\bibitem{Klimov-2008} Klimov A B, Munoz C, Fernandez A, Saavedra C 2008 {\it Phys. Rev. A} {\bf 77} 060303(R).

\bibitem{cripto} Bechmann-Pasquinucci H and Peres A 2000 {\it Phys. Rev. Lett.} {\bf 85} 3313.

\bibitem{Wootters-1987} Wootters W K 1987 {\it Ann. Phys. (N.Y.)} {\bf 176} 1.

\bibitem{telep} Koniorczyk M, Buzek V, and Janszky J 2001 {\it Phys. Rev. A} {\bf 64} 034301.

\bibitem{err} Paz J P, Roncaglia A J, and Saraceno M 2005 {\it Phys. Rev. A} {\bf 72} 012309.

\bibitem{Hayashi} Hayashi A, Horibe M, and Hashimoto T 2005 {\it Phys. Rev. A} {\bf 71} 052331.

\bibitem{Wootters-1989} Wootters W K and Fields B D 1989 {\it
Ann. Phys. (N.Y.)} {\bf 191} 363.

\bibitem{Band} Bandyopadhyay S, Boykin P O, Roychowdhury V, and Vatan V 2002 {\it Algorithmica} {\bf 34} 512.

\bibitem{Zeilinger} Lawrence J, Brukner C, Zeilinger A 2002 {\it Phys. Rev. A.} {\bf 65} 032320.

\bibitem{Wootters-ibm} Wootters W K 2004 {\it IBM J. Res. Dev.} {\bf 48} 99.

\bibitem{Gibbons}Gibbons K S, Hoffman M J, and Wootters W K 2004 {\it Phys. Rev. A} {\bf 70}, 062101.

\bibitem{Klimov-2007} Klimov A B, Romero J L, Bj\"{o}rk G, S\'anchez-Soto L L 2007 {\it J. Phys. A: Math. Theor.} {\bf 40} 3987.

\bibitem{Bjork-2007} Bj\"{o}rk G, Romero J L, Klimov A B, S\'anchez-Soto L L 2007 {\it J. Opt. Soc. Am. B} {\bf 24} 371.

\bibitem{Ghiu} Ghiu I 2012 {\it J. Phys.: Conf. Ser.} {\bf 338} 012008.

\bibitem{Klimov-2009} Klimov A B, Romero J L, Bj\"{o}rk G, S\'anchez-Soto L L 2009 {\it Ann. Phys. (N.Y.)} {\bf 324} 53.


\end{thebibliography}
\end{document}